# Quantitative measurements of biological/chemical concentrations using smartphone cameras

**Zhendong Cao**
EIM Technology
zcao@eimtechnology.com

**Hongji Dai**
EIM Technology
tdai@eimtechnology.com

**Zhida Li**
New York Institute of Technology-Vancouver
zhidali@ieee.org

**Ash Parameswaran**
School of Engineering Science,
Simon Fraser University
paramesw@sfu.ca

**Abstract**

This paper presents a smartphone-based imaging system capable of quantifying the concentration of an assortment of biological/chemical assay samples. The main objective is to construct an image database which characterizes the relationship between color information and concentrations of the biological/chemical assay sample. For this aim, a designated optical setup combined with image processing and data analyzing techniques was implemented. A series of experiments conducted on selected assays, including fluorescein, RNA Mango, homogenized milk and yeast have demonstrated that the proposed system estimates the concentration of fluorescent materials and colloidal mixtures comparable to currently used commercial and laboratory instruments. Furthermore, by utilizing the camera and computational power of smartphones, eventual development can be directed toward extremely compact, inexpensive and portable analysis and diagnostic systems which will allow experiments and tests to be conducted in remote or impoverished areas.

## I. Introduction

In the life sciences, biological samples are collected to measure the concentrations of various compounds of interest, including hormones, proteins, and enzymes. The most common instruments adopted in many laboratories are plate-readers and spectrophotometers. Plate-readers give highly accurate reading of the concentration of fluorescence sample by measuring the ratio of output fluorescent light intensity and input excitation light intensity [1]. Spectrophotometers do the measurements based on absorption approach, which computes the ratio of output absorbed light intensity and incident monochromatic light intensity.

Both instruments are designed to be highly sensitive and accurate, but they sacrifice portability, and their costs are also very high, which make them extremely unlikely to be used in field applications, remote areas or economically impoverished environments. Compared to the standard spectrophotometer or plate reader used in laboratories, a smart phone-based detection and analysis system will have numerous advantages in terms of device portability, cost and ease of operation. The entire system can be extremely compact, inexpensive and battery powered. The smart phone itself can serve as a computing station for data analysis and, thus, a separate computer is no longer needed. With a phone-based detection and analysis system, quantitative measurements of several biological samples can be performed in an easier and cheaper way, allowing researchers to conduct biological experiments in the field or remote areas. Therefore, the development of a phone-based system to quantify compounds of interest and a comprehensive comparison between such a system and existing commercially-available instruments will be the major motivation of this study.

## II. Methodology

When a biological or chemical sample interacts with light, the changing of its concentration may cause a variation of color. A designated smartphone camera based imaging device was built to detect changes in color and map these changes to meaningful physical quantities.

The general functionality and engineering model of the smartphone imaging system is represented in Figure 1.

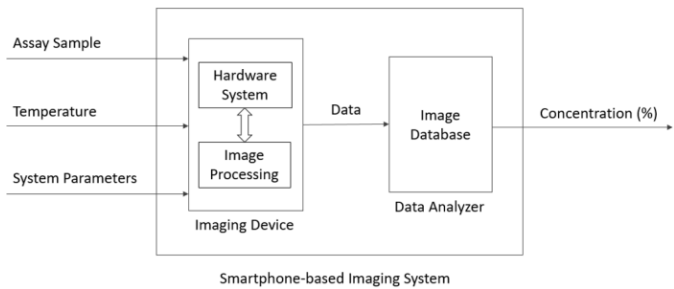

Figure 1: Engineering model of the entire system

In its simplest form, the entire system can be represented as an imaging device and a data analyzer. To quantify the concentration of compounds within a sample, the system takes several input parameters including the assay type, smartphone settings, spectrum of the light source and temperature. These input quantities allow data to be generated from the imaging device, and with a pre-established database, the concentration of the compound within the sample can be estimated.

**a. Assay Preparation**

In this paper, we focus on chemical and biological assays that are in liquid state. Solutions with various concentrations can be obtained using serial dilutions. For all dilutions, water is used as the solvent. Particularly, two types of assay samples are involved in this research:

Type 1: Fluorescent materials, namely:
- Fluorescein
- RNA Mango

Type 2: Liquid colloids, namely:
- Homogenized Milk
- Saccharomyces Cerevisiae

Deep ultraviolet (UV) light will be utilized to excite RNA Mango, therefore quartz micro-cuvettes are used for all the experiments due to their exceptionally wide transmission bandwidth. Figure 2 [2] shows that the transmission of typical quartz's cuvettes is close to 90% for deep UV under 280nm.

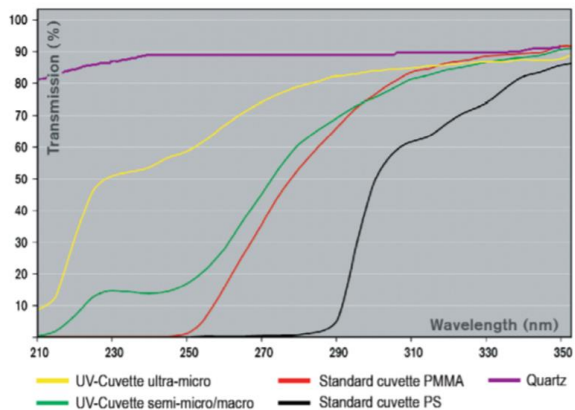

Figure 2 [2]: Transmission curves of cuvettes made from different types of materials

**b. Hardware System**

The entire hardware system of the smartphone-based imaging device can be illustrated in Figure 3 (a). Two identical LEDs are mounted vertically into the two LED holes (left and right) giving an upward illumination towards the samples. This right-angle setup can effectively avoid capturing unexpected light from the LEDs [3], since our target of interest is the output light only. The LEDs were chosen to cover the spectrum to excite the fluorescent assay sample (elaborated in next section). Once the cuvettes are in position, the entire housing will be fully enclosed by the dark box to prevent ambient light leaking inside the enclosure. Users can turn on the device from the electronics module on the backside and align the smartphone camera to the camera hole at the front face for observation and imaging, see Figure 3 (b).

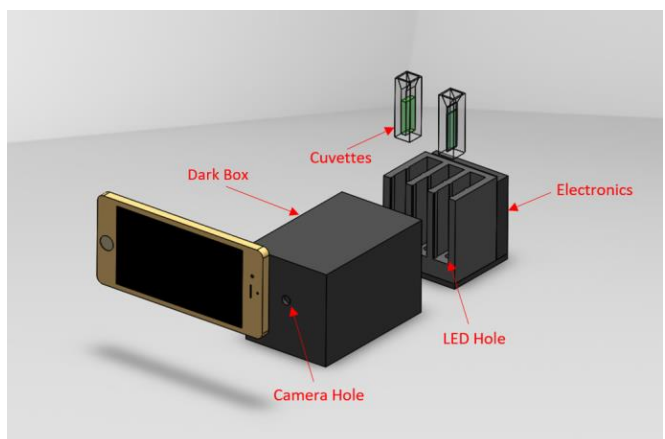

(a): before assembled

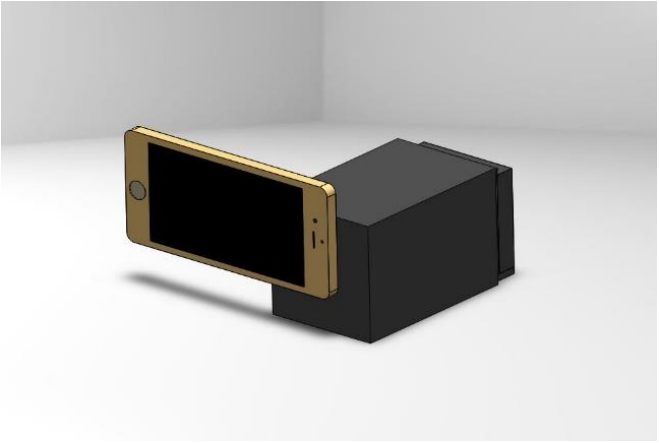

(b): After assembled

Figure 3: CAD model of the enclosure design

## c. Data Collection

Collection of image data is structured as shown in Figure 4. The test is conducted at room temperature. Images are acquired of different sample concentrations at each of the control parameters including LED intensity, smartphone types, exposure time, focus and ISO settings.

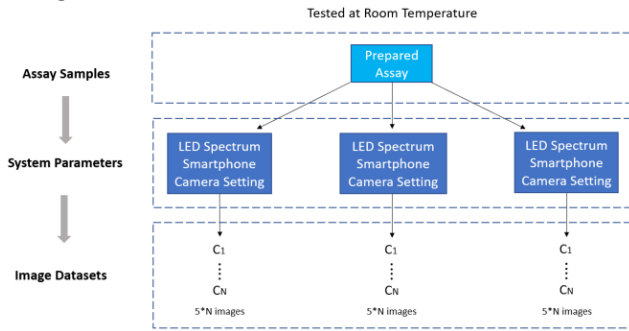

Figure 4: Process of acquiring image data

The exposure time, ISO and aperture, sometimes called the "exposure triangle", used to capture the static image of interest can affect experimental results. ISO controls the gain, or the sensitivity of the camera to the light incident on the image sensor. Exposure time is the duration for which the image sensor stays active to the exposed light. The relationship between pixel value of the captured image and luminance of the object is modeled in equation (1) [4]:

$$N_d = \left(\frac{K_c}{f_s^2}\right)(tS)L_s \qquad (1)$$

where the quantities are

- $K_C$   Calibration constant for the camera
- $t$     Exposure time in seconds
- $f_s$   Aperture number
- $S$     ISO setting in seconds
- $L_S$   Luminance of the scene

The only sources of light in our system are from the LEDs and the fluorescent light emitted by the compound in the sample solution. Given the condition that LED intensity remains constant, the change in luminance, $L_s$, is mainly due to sample emission, which is related to the concentration of the compound of interest. Identical smartphones are used for each set of the experiment, thus the calibration constant and aperture are fixed numbers. Eventually, the only two quantities in this equation that may vary are the ISO setting, S, and exposure time, t. To establish a relationship between image sensor output and fluorescence intensity, ISO and exposure time should be manually adjusted to a set of fixed values for each experimental step.

## d. Image Processing

Image data collected at different conditions were processed using several steps (Figure 5).

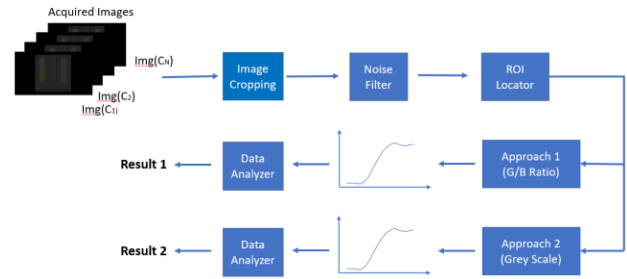

Figure 5: Image processing flowchart

In this stage, image processing was implemented using ImageJ. Repetitive images will be averaged to minimize the noise occurred from the CMOS sensor. The region of interest (ROI) will be manually located on the software and will be analyzed using RGB color model (Figure 6).

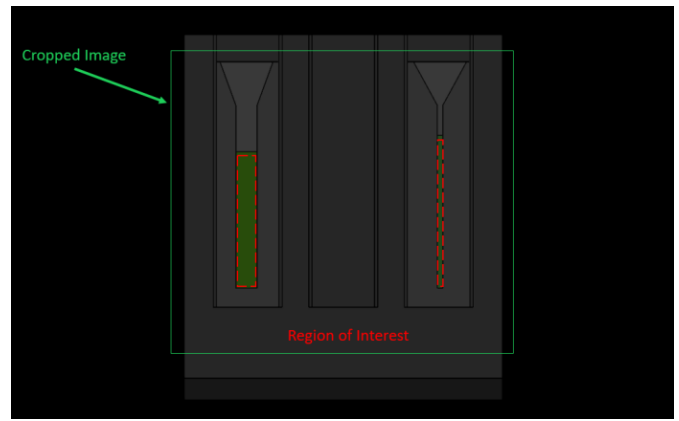

Figure 6: Region of interest from the cropped image

In the specified ROI, the RGB colors are interpreted in approaches for the same image data: 1) the ratio of two channels and 2) grey scale value (average pixel value of three channels).

Mathematically, for a specific ROI with size of M×N pixels, the Approach 1, or the ratio of two channels (e.g. G/B ratio) can be calculated as:

$$\text{G/B Ratio} = \sum_{i=1}^{M}\sum_{j=1}^{N} p(g)_{i,j} \Big/ \sum_{i=1}^{M}\sum_{j=1}^{N} p(b)_{i,j} \quad (2)$$

and the Approach 2, or the grey scale value, can be computed as:

$$\text{Grey Scale Value} = \frac{1}{M \times N}\sum_{i=1}^{M}\sum_{j=1}^{N} \frac{p(r)_{i,j} + p(g)_{i,j} + p(b)_{i,j}}{3} \quad (3)$$

where $p(g)_{i,j}$ and $p(b)_{i,j}$ are the 8-bit green and blue pixel values at location (i, j).

The advantage of the G/B ratio approach is that the measurement is less affected by changes in camera exposure time. If the measurement result is based only on a single channel (e.g. Green), using different exposure times changes the total radiation received by the CMOS sensor, thus causing a shift of channel values. However, when using two color channels, the change in radiation energy received by each color channel can be effectively cancelled when their ratio is considered [5].

Alternatively, the grey scale approach sums up the intensities of all color channels of the smartphone to determine the output value, thus it is more sensitive to detect the intensity change of visible light. The drawback of grey scale approach is that it only tells the change of light intensity but provides no information about the phase of the spectrum.

**e. Data Analyzer**

Measurement approaches with both G/B ratio and grey scale should exhibit consistency in terms of intensity measurement. By plotting the response curves at different concentrations, we can determine the quantitative results for each measurement. These results can characterize the performance of our system, including important parameters such as detection limit, sensitivity, repeatability and accuracy. Figure 7 illustrates these definitions pictorially.

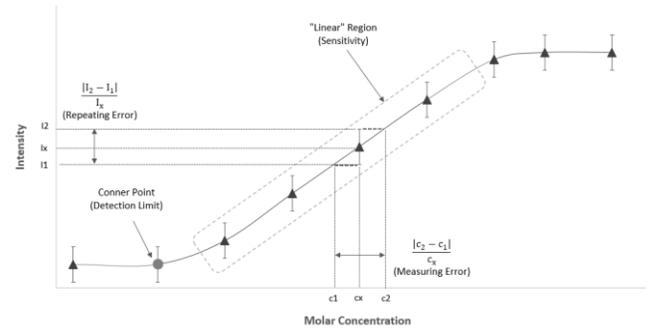

Figure 7: Illustration of repetition error, sensitivity, measurement error and detection limit

Detection limit refers to the lowest concentration at which the fluorescence signal can be picked up and quantified by the system, and it can be identified by the change in slope in the response curve.

To quantify sensitivity, we use a linear curve fitting method to model the function mathematically in the region where system output response changes consistently in response to the concentration changes. For consistency, the curve fitted line should have an R-square value of at least 0.95. The slope or gradient of the curve in the sensitive region is defined as sensitivity.

The repeating error indicating the errors for the system responses obtained from successive measurements under the same test conditions is defined as:

$$\text{Repeating Error} = \left(\frac{|I_2 - I_1|}{I_x}\right) \times 100\% \quad (4)$$

where $I_x$ is the intensity of $c_x$, $I_1$ and $I_2$ are intensities corresponding to the lower and higher concentrations respectively. This quantity can be represented by the vertical error bar on the response curve in Figure 7.

The measurement error is defined as the error between the result obtained by our device and the true physical value of the assay concentration. Mathematically, measurement error is expressed as:

$$\text{Measuring Error} = \left(\frac{|c_2 - c_1|}{c_x}\right) \times 100\% \quad (5)$$

where $c_1$ and $c_2$ are the two concentration values corresponding to the lower error bar and higher error bar of the point being measured. For a measurement to be accurate, the value of accuracy should be small. For example, $(10 \pm 5\%)$ nM is relatively more accurate than

$(10 \pm 20\%)$ nM. Since the error bars for each sample point may vary, we will accommodate the worse case result as the overall resolution of the entire system.

**f. Image Database**

Implementation of our quantification system requires a database that converts data from the imaging device to concentrations of the sample. The imaging device and the data acquisition system generate a series of output data containing:

- Assay information
- Temperature
- Smartphone, LED power, Exposure Time, ISO settings
- Channel Ratio Value, grey scale value

The system takes the above information and outputs the closest concentration determined by using the database. This database is established from the test results conducted for known sample concentrations at given conditions. For an actual experiment when the sample concentration is unknown, the system will perform reverse mapping based on the pre-established test results. The accuracy and precision of the measurement can be improved by repetitively collecting a large volume of test results.

### iii. Results

**a. Fluorescein**

Fluorescein is an organic chemical compound that is commonly used as fluorescent dye. Fluorescein has high molar absorptivity at 480 nm, high fluorescence quantum yield and good water solubility, which makes it very useful for detection and optical imaging [6]. To setup a test for fluorescein, we chose cyan LEDs with 477nm peak wavelength as excitation light sources. Figure 8 demonstrates the setup of experiment.

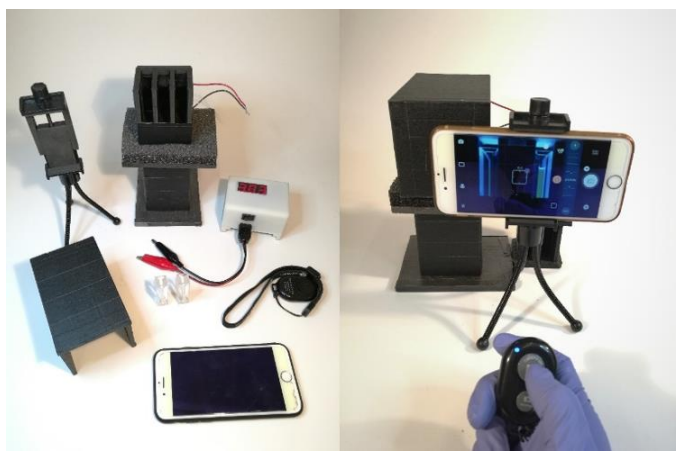

Figure 8: Hardware setup for smartphone based fluorescence measurement

Once images are collected, they are processed with ImageJ. By manually locating the ROI on the image, the software computes the average RGB values of the selected ROI, as shown in Figure 9.

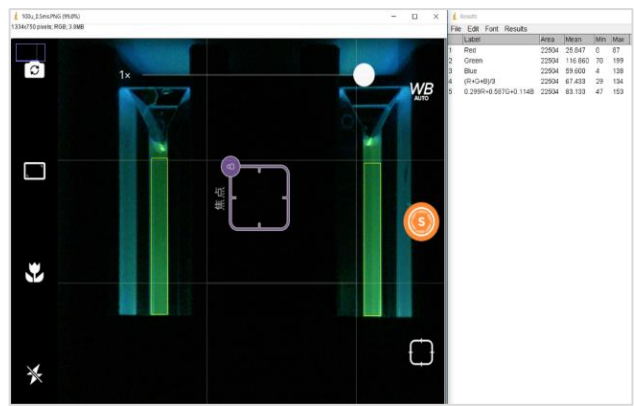

Figure 9: Interface of ImageJ showing the selection of ROI and calculation of the average pixel values in this region

The RGB values acquired from each ROI will be interpreted using the two approaches (G/B ratio and Grey Scale). The data obtained at each corresponding assay concentration will be used to generate plots, as shown in Figure 10 and Figure 11.

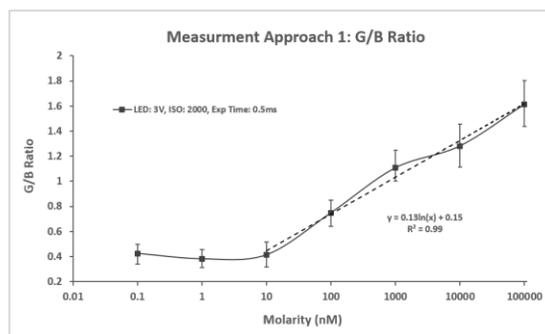

Figure 10: G/B ratio vs. molarity at 22°C

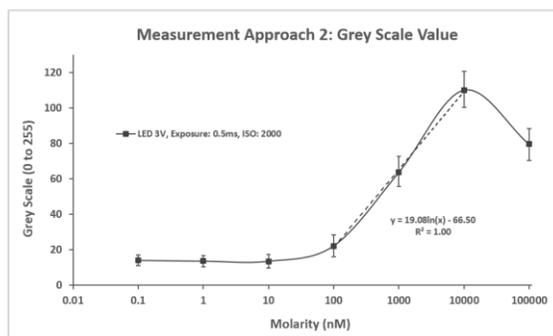

Figure 11: Gray scale value vs. molarity at 22°C

One observation is that the output response drops at 100uM for approach 2 while it is still increasing in approach 1. This phenomenon can be explained by the inner filter effects of fluorophores caused by the increasing possibility of collisional quenching at high concentrations. Under inner filter effects, output fluorescence intensity is weaken therefore causing a drop in Figure 11, however the weakened intensity may not be picked by using approach 1 since approach 1 is more sensitive to phase shift instead of intensity. To avoid confusion and further complexity involved with detecting high concentrated samples, we set the upper limit detection range of our system to 10uM.

The parameters representing the characteristics of the smartphone imaging system with the two measurement approaches are listed in Table 1.

**Table 1: Results summary of two measurement approaches (Fluorescein)**

|  | G/B Ratio | Grey Scale |
|---|---|---|
| Detection Limit | 10nM-10uM | 10nM-10uM |
| Max Repeating Error | 42.8% | 56.1% |
| Average Repeating Error | 29% | 38% |
| Sensitive Range* | 10nM – 10uM | 100nM – 10uM |
| Sensitivity | 0.13/10-fold | 19.08/10-fold |

In the "linear" region of the plots where the slopes are approximated by the curve fitting logarithm functions, we derived the inverse functions of the curve fitted functions. Using Equation (6), (7), we can approximate the concentration of the assay using either G/B ratio or grey scale approach:

$$c_{G/B} = 0.315 e^{7.69(x)} \qquad (1)$$

$$c_{Grey} = 32.63 e^{0.0524(x)} \qquad (7)$$

where x is the system output value measured by G/B ratio or grey scale approach. By comparing the characteristic curves of the two systems above, Equation 7 (the Grey Scale approach) gives a more precise mapping from intensity to concentration than Equation 6 (the G/B approach).

**b. RNA Mango**

RNA Mango is a high affinity RNA aptamer that amplifies the fluorescence signal of RNA, allowing the purification of fluorescently tagged RNA complexes to be tracked in real time [7]. RNA Mango has two absorption maxima located at 260 nm and 510 nm. The maximum emission wavelength of RNA mango is 535 nm, which is a green dominant color. Since the lowest detectable wavelength for both human eyes and CMOS image sensors in smartphones is around 400 nm [8], using a deep UV light source to excite RNA Mango can effectively minimize the interference to our smartphone based imaging system. The hardware setup is demonstrated in Figure 12.

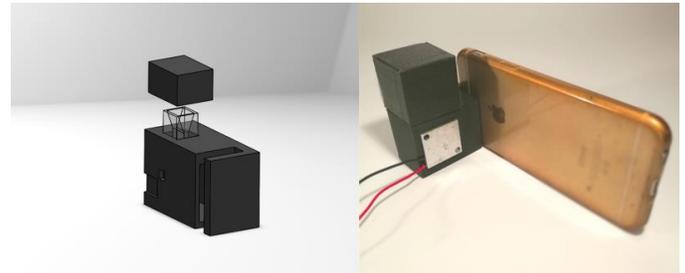

Figure 12: Hardware setup for RNA Mango experiment

Following the same procedures to fluorescein test, the results for RNA Mango concentrations measured with G/B ratio and grey scale are plotted in Figure 13 and Figure 14.

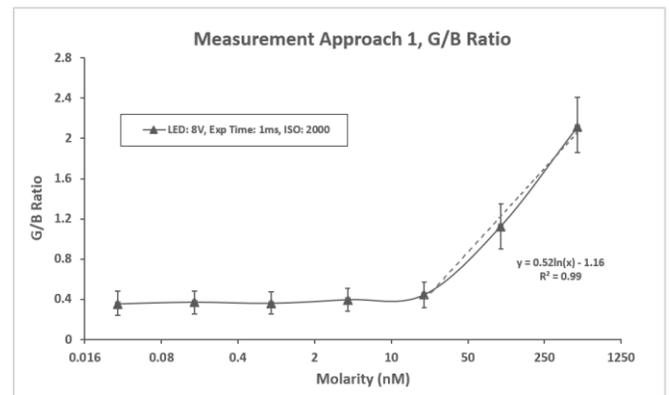

Figure 13: G/B ratio vs. molarity measured at 22°C (RNA Mango)

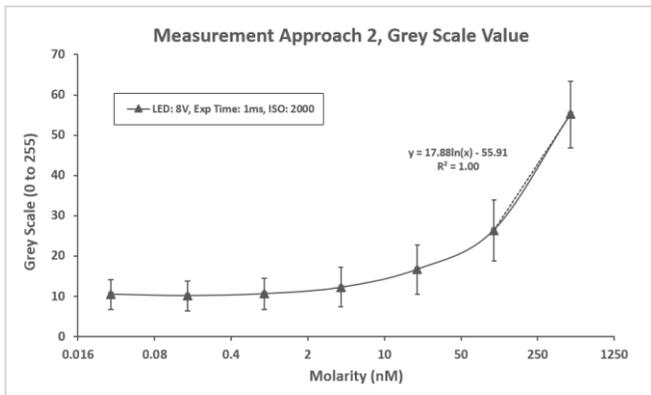

Figure 14: Grey scale value vs. molarity measured at 22°C (RNA Mango)

The system performance of the two measurement approaches is listed in Table 2.

Table 2: Results summary of two measurement approaches (RNA Mango)

|  | G/B Ratio | Grey Scale |
|---|---|---|
| Detection Limit | 20 nM | 20 nM |
| Max Repeating Error | 68.4% | 75.3% |
| Average Repeating Error | 53% | 65% |
| Sensitive Range | > 20 nM | > 100 nM |
| Sensitivity | 0.52/5-fold | 17.88/5-fold |

**c. Homogenized Milk**

Generally, absorption and scattering may occur when light interacts with a colloidal solution. For scattering based measurements, output light is caused by physical collisions between photons and particles in the liquid, thus, the spectrum of the output light is the same as that of the input beam. However, changing of concentrations will affect the probability of collisions among photons and particles, causing the change of output light intensity where the color change can be detected [9], as shown in Figure 15.

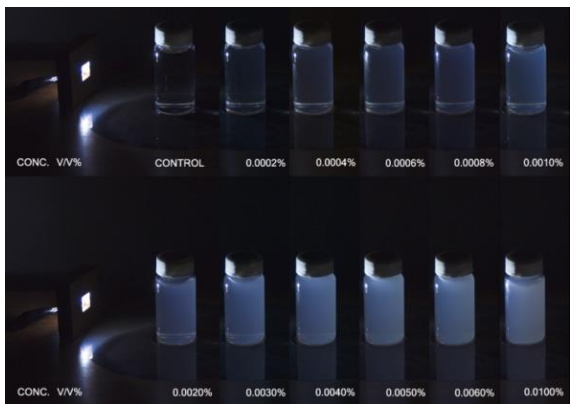

Figure 15 [10]: Scattering effects of milk at different concentrations

Homogenized milk is a colloidal solution, more specifically, an oil-in-water emulsion that contains fat globules and other soluble substances such as proteins, lactose, vitamins and minerals. We chose a skimmed milk (2% fat) as the sample to be tested, which corresponds to 20g/L of fat based on the nutrition facts provided by the producer. The fat percentage of the milk sample was serially diluted using a 5-fold dilution. The results are plotted in Figure 16 and Figure 17.

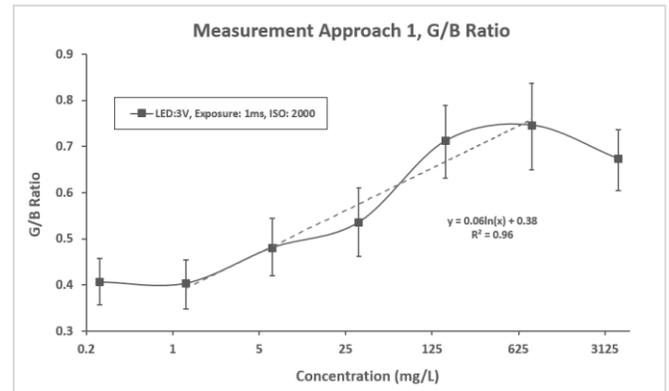

Figure 16: G/B ratio vs fat concentrations of milk, measured at 22°C (2% Milk)

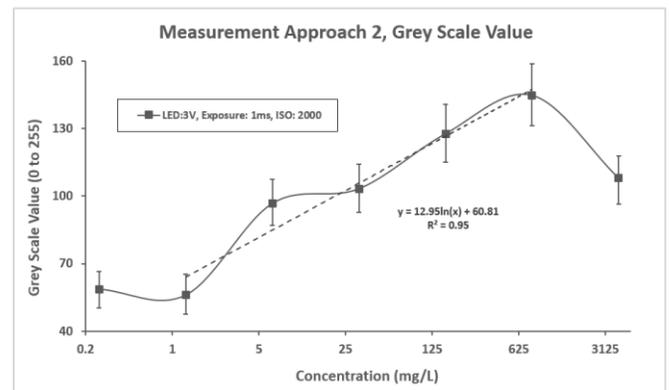

Figure 17: Grey scale value vs fat concentration of milk, measured at 22°C (2% Milk)

Table 3 tabulates the system parameters with two measurement approaches.

Table 3: Results compared by two measurement approaches (2% Milk)

|  | G/B Ratio | Grey Scale |
|---|---|---|
| Detection Limit | 1.28-800 mg/L | 1.28-800 mg/L |
| Max Repeating Error | 27.7% | 27.8% |
| Average Repeating Error | 24.9% | 25.4% |
| Sensitive Range* | 1.28 – 800 mg/L | 1.28 – 800 mg/L |
| Sensitivity | 0.06/5-fold | 12.95/5-fold |

### d. Saccharomyces Cerevisiae

Saccharomyces cerevisiae (yeast) are used as another colloidal test sample. Given the growth characteristics of yeast cells in a suitable growth media, the concentration being measured generally does not remain constant across time. Replication of yeast cells can be dramatically fast in log phase which takes place around 1 hour after cells are put in growth media [11]. Measuring in yeast cell concentrations while cells are growing fast should be avoided, so each measurement must be conducted immediately to minimize cells replication time. The colloidal of yeast cells are prepared by mixing dry yeast powder with deionized distilled water. Serial dilution was applied to obtain different concentrations. Results are shown in Figure 18 and Figure 19 respectively.

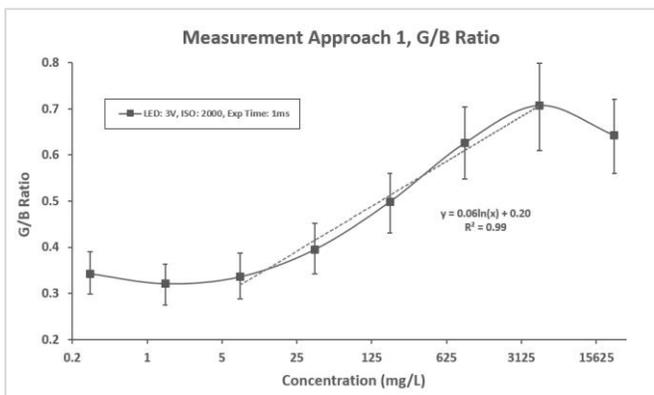

Figure 18: G/B ratio vs. concentration, measured at 22°C (Yeast)

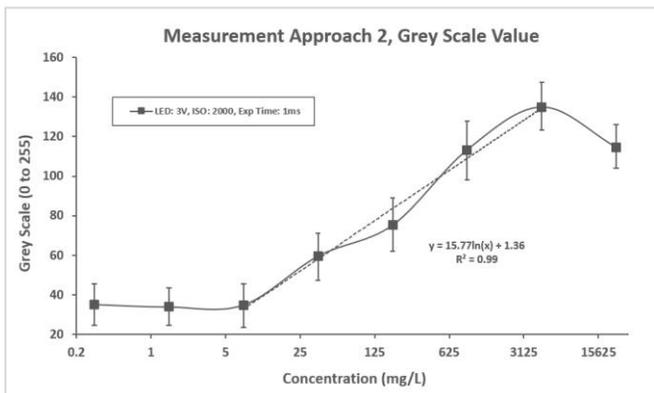

Figure 19: Grey scale value vs. concentration, measured at 22°C (Yeast)

The upper and lower detection limit and other parameters such as sensitivity and repetition errors can be quantified in Table 4.

**Table 4: Results compared by two measurement approaches (Yeast)**

|  | G/B Ratio | Grey Scale |
|---|---|---|
| Detection Limit | 7.36-4600 mg/L | 7.36-4600 mg/L |
| Max Repeating Error | 39.4% | 29.3% |
| Average Repeating Error | 27.8% | 31.6% |
| Sensitive Range* | 7.36 – 4600 mg/L | 7.36 – 4600 mg/L |
| Sensitivity | 0.06/5-fold | 15.77/5-fold |

### VI. System Validation

The results obtained from our system will be compared with commercial laboratory instruments for validation purpose. The instruments used for comparison are fluorescence based microplate reader (fluorescein), spectrophotomer (RNA Mango) and high accuracy optical powermeter (milk and yeast). The output data are normalized such that all system output reading values are scaled within range of [0,1] therefore response curves of different systems can be merged in a single plot for better visualization. Figure 20-23 give the graphical comparison between our system performance to the commercial instruments.

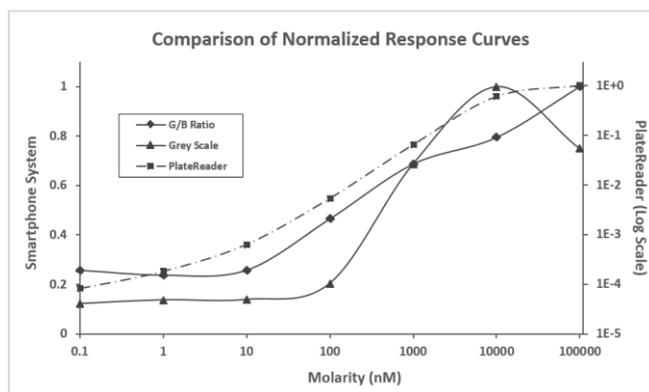

Figure 20: Comparison of response curves of the smartphone based system and commerical platereader

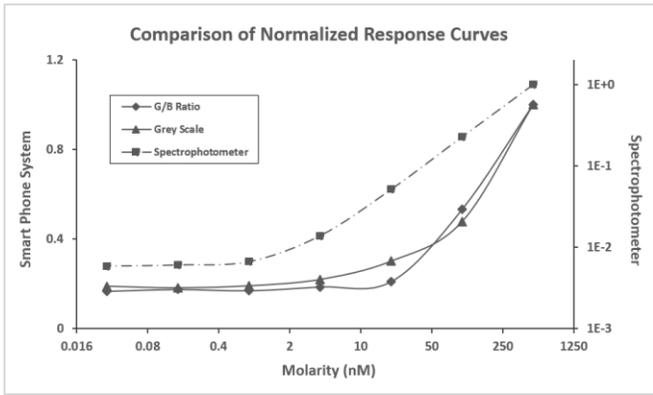

Figure 21: Comparison of response curves of the smartphone based system and spectrophotometer

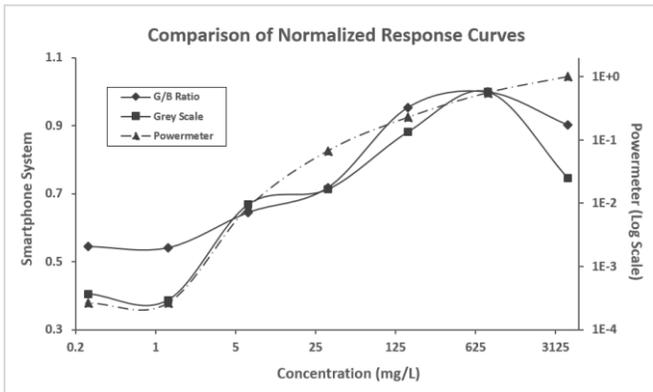

Figure 22: Comparison of response curves of the smartphone based system and power meter (fat concentration in homogenized milk)

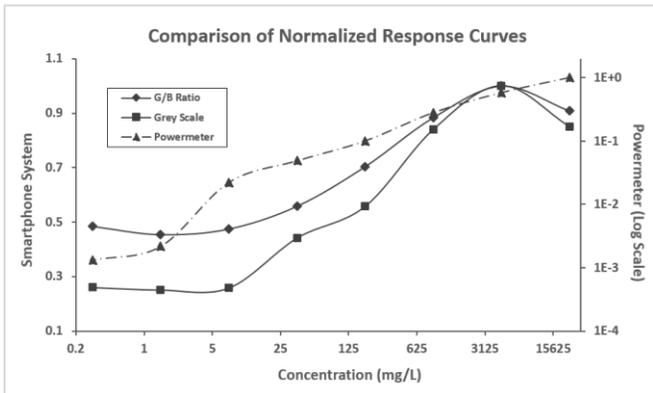

Figure 23: Comparison of response curves of the smartphone based system and power meter (yeast)

## V. Conclusion

In this project we explored the concept of using smartphones as the sensor to detect and quantify the concentration of several biological samples. Our studies focused on fluorescent materials and colloidal mixtures. A portable and compact hardware system was built to accomplish the following essential requirements of the experiment:

- Provide secure housing for standard cuvettes
- Eliminate light interference from the external environment
- Create a right-angle layout for excitation and emission light
- Fix the position of the observing smartphone

Procedures for image acquisition and data processing were also established for the system to generate precise and consistent results throughout experiments. From the collected measurement results of each experiment, we constructed a database which allows the estimation of chemical concentrations of samples, interpreted from acquired image data.

To compare the performance of our system, all experiments conducted were also replicated using commercial instruments such as microplate readers, spectrophotomers and power meters. As expected, commercial instruments demonstrated better performance, particularly with respect to the detection limit and sensitivity of concentration changes. However, the smartphone-based imaging system exhibited relatively good ranges of detections limits and sensitivities and performed better than commercial instruments with regards to its repeatability. Our demonstrated device is small, portable and inexpensive, which makes it remarkably suitable for quick tests and experiments conducted in areas where electrical power is unreliable or not available.